\documentclass[12pt]{article}
\usepackage{amssymb,latexsym}
\oddsidemargin=\evensidemargin
\advance\topmargin by-\headheight
\def\A{{\cal A}}

\def\CC{{\mathbb C}}

\def\F{{\mathbb F}}
\def\FF{{\cal F}}

\def\GL{\mathop{\rm GL}\nolimits}
\def\GLZ{\mathop{\rm GL}\nolimits_2({\mathbb Z})}
\def\gm{\Gamma}
\def\id{\mathop{\rm id}\nolimits}
\def\im{\mathop{\rm im}\nolimits}
\def\ker{\mathop{\rm ker}\nolimits}
\def\lg{\langle}

\def\Mat{\mathop{\rm Mat}\nolimits}
\def\mapright#1{\smash{\mathop{\longrightarrow}\limits^{#1}}}

\def\ov{\overline}
\def\P{{\cal P }}
\def\PGL{\mathop{\rm PGL}\nolimits}

\def\R{{\cal R}}
\def\rg{\rangle}
\def\S{{\cal S}}
\def\SLZ{\mathop{\rm SL}\nolimits_2({\mathbb Z})}

\def\X{{\cal X}}

\def\ZZ{{\mathbb Z}}

\def\proof{{\bf Proof}.\ }
\def\bull{\vrule height .9ex width .8ex depth -.1ex }
\def\subs{\stepcounter{subsection}{\bf\thesubsection{.}}
\addtocounter{subsection}{-1}\refstepcounter{subsection}}
\def\subsn{\vspace{2mm}\subs}

\newtheorem{formula}{}[section]

\newtheorem{definition}[formula]{Definition}
\newtheorem{corollary}[formula]{Corollary}
\newtheorem{remark}[formula]{Remark}
\newtheorem{lemma}[formula]{Lemma}
\newtheorem{theorem}[formula]{Theorem}
\newtheorem{problem}[formula]{Problem}

\def\thrm{\begin{theorem}}
\def\thrml#1{\begin{theorem}\label{#1}}
\def\ethrm{\end{theorem}}
\def\rmrk{\begin{remark}}
\def\rmrkl#1{\begin{remark}\label{#1}}
\def\ermrk{\end{remark}}
\def\dfntn{\begin{definition}}
\def\dfntnl#1{\begin{definition}\label{#1}}
\def\edfntn{\end{definition}}
\def\nmrt{\begin{enumerate}}
\def\enmrt{\end{enumerate}}

\def\prblm{\begin{problem}}
\def\prblml#1{\begin{problem}\label{#1}}
\def\eprblm{\end{problem}}
\def\qtn{\begin{equation}}
\def\qtnl#1{\begin{equation}\label{#1}}
\def\eqtn{\end{equation}}
\def\lmm{\begin{lemma}}
\def\lmml#1{\begin{lemma}\label{#1}}
\def\elmm{\end{lemma}}
\def\crllr{\begin{corollary}}
\def\crllrl#1{\begin{corollary}\label{#1}}
\def\ecrllr{\end{corollary}}

\begin{document}
\title{
Homomorphic public-key cryptosystems \\
over groups and rings}
\author{
Dima Grigoriev \\[-1pt]
\small IRMAR, Universit\'e de Rennes \\[-3pt]
\small Beaulieu, 35042, Rennes, France\\[-3pt]
{\tt \small dima@math.univ-rennes1.fr}\\[-3pt]
\small http://name.math.univ-rennes1.fr/\~{}dimitri.grigoriev
\and
Ilia Ponomarenko\\[-1pt]
\small Steklov Institute of Mathematics,\\[-3pt]
\small Fontanka 27, St. Petersburg 191011, Russia\\[-3pt]
{\tt \small inp@pdmi.ras.ru}\\[-3pt]
\small http://www.pdmi.ras.ru/\~{}inp
\thanks{Partially supported by RFFI, grants,  03-01-00349,
NSH-2251.2003.1 and a grant of NATO. The author would like to thank the
Mathematical Institute of the University of Rennes during the stay in which
this paper was initiated.
}
}
\date{31.08.2003}
\maketitle

\begin{abstract}
We propose a new homomorphic public-key cryptosystem over arbitrary
nonidentity finite group based on the difficulty of the membership
problem for groups of integer matrices. Besides, a homomorphic cryptosystem
is designed for the first time over finite commutative rings.
\end{abstract}

\section{Introduction}

\subs
The problem of constructing  reliable cryptosystems for secret computations
had been extensively studied last years (see~\cite{B,BY88,FM,GP,RAD}).
Generally, it consists in encryption of a circuit over an algebraic structure
$H$ (e.g. group, ring, etc.). One of possible approaches to it is to find a
publically known algebraic structure $G$ and a secret homomorphism $f:G\to H$.
If the inversion of~$f$ is efficiently computable and computing of $f$ is a
hard computational problem (i.e. $f$ is a trapdoor function), one can design
a homomorphic public-key cryptosystem in which an element $h\in H$ is encrypted
by an element of the form $gg_h$ where $g$ is a random element of $\ker(f)$
and $f(g_h)=h$.
Using such a cryptosystem one can efficiently implement a secret computation
given by any circuit over the structure~$H$. Some other applications of
homomorphic public-key cryptosystems can be found in~\cite{B,D02,DC97,ST98}.
We mention also that the group theory is a source of constructions (apart from
homomorphic cryptosystems) in the cryptography, see e.g. \cite{Gr,K,MW,NS,PKHK}.

It is well known that any boolean circuit of logarithmic depth can be
efficiently simulated by a circuit over an arbitrary finite nonsolvable group,
see~\cite{Ba} (another approach to encrypting boolean circuits was undertaken
in~\cite{SYY}). Thus one of the first natural problems concerning
secret computations is to design a homomorphic public-key cryptosystem over
a finite group. The known examples of such systems include the quadratic residue
cryptosystem (see~\cite{GM,GB}) over the group of order~2 and the
cryptosystems (see~\cite{OU98,P99,Ra}) over some cyclic and dihedral groups.
However, in these and some
other cryptosystems the involved groups are solvable and so can not be
used for the above cited simulation of boolean circuits. The first
homomorphic public-key cryptosystem over an arbitrary nonidentity finite group
was designed in~\cite{GP}.

It should be mentioned that the secrecy of all
these cryptosystems was based on the difficulty of some problems closely
related with that of integer factoring. However, ``as long as factoring
remains intractable, we are in a good position, but we are overindependent on
the computational complexity of one particular problem'' \cite{WM85}. In
addition, unlike factoring it is unknown whether there is a quantum machine
which can decide the membership to a non-abelian matrix group, the problem
on which relies the security of the cryptosystems in the present paper.
In contrast to the cryptosystems based on the factoring problem the first main
result of this paper is a new homomorphic public-key cryptosystem over
arbitrary nonidentity finite group based on the difficulty of the membership
problem for groups of integer matrices (for details see Section~\ref{120403a}
and Theorem~\ref{110303b}).

\thrml{00}
For a nonidentity finite group $H$ given by generators and relations one can
choose a group $G\le\GLZ$ and a homomorphism $f:G\to H$ to obtain a
homomorphic public-key cryptosystem over~$H$.
\ethrm

We may think of $H$ to be a finite small group. On the other hand, the
infiniteness of $G$ is not an obstacle for performing algorithms of encrypting and
decrypting (for the latter using the trapdoor information) since they involve
just calculations with integer $2\times 2$ matrices. In this connection we mention a public-key cryptosystem
from  \cite{DJSS} in which $f$ was the natural epimorphism from a free group $G$ onto
the group $H$   given by generators
and relations. In this case for any element of $H$ one can produce its preimages
(encryptions) by inserting in a word (being already a produced preimage of $f$)
from $G$ any relation defining $H$. In other terms, decrypting of $f$ reduces to
the word problem in $H$. In our approach the epimorphism $f$ is given on
specially chosen generators of an appropriate subgroup $G$ of a free group
$F\subset\GLZ$, and the trapdoor consists in a
polynomial-time algorithm (see Subsection~\ref{110303a}) which allows one to
represent an element of $G$ (being
an integer matrix) as a product of free generators of~$F$.
Publically in the cryptosystem from Theorem~\ref{00} a certain set of
generators of $G$ is exhibited, and the security of the cryptosystem
relies on the difficulty (without knowledge of the trapdoor) of finding a
representation of an element of $G$ as a product of these generators,
while in~ \cite{DJSS} an element of the free group G is given just by
means of a product of its generators. (In fact, we keep a secret ``good''
basis of~$F$ which enables us to compute
matrices of $G$ easily; at the same time the public key is given
by a ``bad'' basis of~$G$ for which the representation problem is supposedly
hard.)

We mention also that two public-key cryptosystems (being not homomorphic)
based on the group $\SLZ$ were suggested in \cite{Y98,Y99} which were
subsequently broken in \cite{S2000,BG}. These cryptosystems were hiding the
generators of a subgroup of $\SLZ$ by means of conjugating them with a
secret matrix.

In \cite{WM85,GZ91} two constructions of cryptosystems (being not homomorphic)
were proposed with the difficulty of breaking relied on the word problem (in
finitely generated groups). The common feature of both papers is that a
public key is given by two words $m_0, m_1$ and a family $\R$ of words.
Then encrypting of a bit $i\in\{0,1\}$ is carried out by means of starting with $m_i$
and subsequent random inserting polynomial number of times of the words from
$\R$. Denote by $G$ the group given by the relations $\R$. Then
basically the trapdoor needs a solution of the word problem in $G$. To this
end the epimorphisms of the form $f:G\rightarrow H$, provided that
$f(m_0)\neq f(m_1)$ were suggested such that the word problem in the group
$H$ is easy, thereby this epimorphism plays a role of a trapdoor. In
\cite{WM85} the epimorphism $f$ consists actually in adding some relations of commutativity
of the generators. In \cite{GZ91} as a group $H$ is taken the
Grigorchuk group with 4 generators (and being not finitely presentable)
corresponding to a certain fast computable infinite word $\chi$. It is shown
in \cite{GZ91} that the word problem in this group is easy, thus $\chi$ plays a
role of a trapdoor. So, the principal difference of the cryptosystems
proposed in \cite{DJSS,WM85,GZ91} from our cryptosystem is that they
perform calculations with words, whereas our cryptosystem deals with integer
$2\times 2$ matrices.

It seems to be an interesting open question whether for a non-abelian group $H$
there exists a homomorphic cryptosystem with a finite group $G$?

\subsn
The second topic of this paper is devoted to homomorphic public-key
cryptosystems over finite rings. This problem was first posed in \cite{RAD}
(see also \cite{FM}) and in \cite{BY88} it was demonstrated that a direct
approach
to it fails. At present there are only a few results in this direction. In
particular, we mention the cryptosystem
from~\cite{D96} based on a homomorphism from the direct sum of rings
isomorphic~$\ZZ$. A finite version of this system \cite{D02} was recently broken
in~\cite{Bao}. As the second main result of this paper we present a
homomorphic public-key cryptosystem over a finite commutative ring (for
details see Section~\ref{sec3}). Before formulating it we recall that any
finite commutative ring with identity is isomorphic to a direct sum of local
rings (see~\cite{MD74}).

\thrml{100303a}
Let $R$ be a finite commutative ring with identity different from a direct sum
of several copies of rings isomorphic to $\ZZ_2$. Then there exists a homomorphic
public-key cryptosystem over $R$ with respect to a homomorphism $f:\A\to R$
for an appropriate finite commutative ring~$\A$.
\ethrm

In the cryptosystem of Theorem~\ref{100303a} the ring $\A$ is a group ring
of a finite Abelian group $G$ and $f$ is the epimorphism induced by a suitable
secret epimorphism from $G$ to the multiplicative group of~$R$. The only
commutative rings for which any homomorphism of such kind is trivial,
have trivial multiplicative groups, and so are the direct sums of copies of
the ring~$\ZZ_2$. Thus the natural open question is to
find a homomorphic public-key cryptosystem over the ring~$\ZZ_2$. The way
we construct the ring~$\A$ gives a  bound on the
cardinality of $\A$ being {\it double exponential} in the cardinality of
$R$. This condition is essential in the following
sense. As we will see in Section~\ref{sec3} any finite ring
of exponential cardinality is a subring of the  ring
$\Mat(n,\ZZ_m)$ of $n\times n$ matrices over $\ZZ_m$ with $n$ and $\log m$
bounded by polynomials. The latter construction of embedding a
ring into a matrix ring
is not efficient a priori, in fact, its efficiency depends on the way in
which the ring is given. On the other
hand, Theorem~\ref{100303e} states that the homomorphisms of the rings
given as subrings of $\Mat(n,\ZZ_m)$ can not be secret.

It should be remarked that secret homomorphisms from Theorem~\ref{100303a}
can not be used for encrypting circuits over rings due to its size. The
problem of finding cryptosystems suitable for such encrypting as well as
constructing secret homomorphisms over noncommutative finite rings are
still open. Theorem~\ref{100303e} shows that if there exists a homomorphic public-key cryptosystem
over a finite ring $R$ with the cardinality of the ring $\A$ being exponential in
the cardinality of $R$ , it should avoid {\it explicit} representing of $\A$ as a
subring of some matrix ring $\Mat(n,\ZZ_m)$.

\section{A homomorphic cryptosystem over a finite group}
\label{120403a}

Throughout the section for a finite set $X$ we denote by $W_X$ the set of all
the words in the alphabet $X^\pm=X\cup X^{-1}$. A word from $W_X$ with no
subword $xx^{-1}$, $x\in X^\pm$, is called irreducible. For an integer
$a\in\ZZ$ we denote by $l(a)$ the bit size of it; for $S\subset\ZZ$ we
set $l(S)=\sum_{a\in S}l(a)$.

\subs\label{110303g}{\bf Representation problem.}
Let $\gm$ be a group and $X$ be a finite subset of~$\gm$. We are interested
in the problem of finding an {\it $X$-representation} of an element $g\in G$
where $G=\lg X\rg$ is a subgroup of~$\gm$ generated by~$X$. By an
$X$-representation of $g$ we mean an irreducible word $w_g\in W_X$ such that
$\pi(w_g)=g$ where $\pi$ is the epimorphism of the free group on~$X$ onto the
group $G$ with $\pi|_X=\id$. Obviously, if $\gm$ is a free group on~$X$,
then $G=\gm$ and each element of~$\gm$ has the unique $X$-representation.
If $w_g=x_1^{a_1}\cdots x_m^{a_m}$ where $x_i\in X$ and $a_i\in\ZZ$ for
all~$i$, then the number $l(w_g)=\sum_il(a_i)$ is called the
{\it bit size} of the $X$-representation $w_g$ of~$g$. We observe that the
size of~$g$ as an element of the group~$\gm$ depending
essentially on the nature of $\gm$ can
substantially differ from the bit size of an $X$-representation of it as well
as the bit sizes of two different $X$-representations of~$g$. In what follows
we look for the algorithms finding $X$-representations of~$g$ efficiently,
i.e. in polynomial time in size of~$g$ in~$\gm$ and in minimal bit size of
its $X$-representation.
\vspace{2mm}

{\bf Representation Problem $\P(\gm,X)$.} Let $\gm$ be a group and
$X\subset\gm$ be a finite set. Given $g\in\lg X\rg$ presented as an
element of~$\gm$ find an $X$-representation of~$g$ efficiently.\bull
\vspace{2mm}

It should be mentioned that the representation problem consists in finding
a certificate for the membership problem when the group in question is
given by generators. If $\gm$ is a symmetric group of degree~$n$, then both
of these problems can be solved in time $n^{O(1)}$ by the sift algorithm (see
e.g.~\cite{L}). However, if $\gm=\GL_n(\ZZ_m)$ then both of these problem are
closely related with the discrete logarithm problem (when $n=1$, $m$ is a
prime and $X$ consists of a generator of the multiplicative group of the
ring~$\ZZ_m$). The representation
problem is NP-hard {\it in average} in general even if $\gm$ is a free group
of a finite rank \cite{W}.

To adapt the representation problem to constructing public-key cryptosystems
we have to describe a trapdoor information providing a polynomial-time
solution of this problem. A general idea can be explained as follows. Let
$G<F<\gm$ be groups and $F=\lg X'\rg$, $G=\lg X\rg$ for
some finite sets $X, X'\subset\gm$. Suppose that both of the problems
$\P(\gm,X')$ and $\P(F,X)$
can be solved efficiently. Then the problem $\P(\gm,X)$ can also be solved
within the same time whenever using the corresponding algorithms one can
find an $X'$-representation and an $X$-representation of an element from
$\lg X\rg$ the bit sizes of which are approximately the same.
In this case one could use the set $X'$ as a trapdoor for the problem
$\P(\gm,X)$.
In the next subsection we realize this idea for $\gm=\GLZ$ and apply it
for constructing a homomorphic public-key cryptosystem over any nonidentity
group given by generators and relations.

\subsn{\bf The main construction.}
Let us define a family of free subgroups of the group $\GLZ$. First we recall
that given an integer $n\ge 2$ the matrices
\qtnl{120403b}
A_n=\left(\matrix{1 & n\cr 0 & 1\cr}\right),\quad
B_n=\left(\matrix{1 & 0\cr n & 1\cr}\right)
\eqtn
form a basis of a free subgroup of the group $\GLZ$ (see \cite[p.232]{LS}).
Next, from the proof of~\cite[Proposition~3.1]{LS} it follows that given
a nonempty set $S\subset\ZZ$ the set
$$
X(n,S)=\{A_n^{-s}B_nA_n^s:\ s\in S\}
$$
is also a basis of a free group $G(n,S)\subset\GLZ$. The following statement
proved in Subsection~\ref{110303a} enables us  to define a homomorphic
public-key cryptosystem with these groups.

\thrml{110303b}
Given an integer $n\ge 2$ and a finite set $S\subset\ZZ$ one can find
the $X(n,s)$-representation $w_g$ of an arbitrary matrix $g\in G(n,S)$
in polynomial time in $l(n)+l(S)+l(w_g)$.
\ethrm

Let $H=\lg\X;\R\rg$ be a nontrivial group given by the set $\X$ of at least
two~\footnote{This is rather technical restriction because even $H$ is a
cycle group one can choose as $\X$ nonminimal set of generators.} generators
and the set $\R$ of relations. Choose randomly $n\ge 2$, sets $S\subset\ZZ$,
$R\subset W_\R$ such that $|S|=|R|=|\X|$, and bijections $h\mapsto x_h$,
$h\mapsto r_h$ from $\X$ to $X(n,S)$ and to $R$ respectively. Set
$$
X=X(n,S,R)=\{x_hr_h:\ h\in\X\},\quad G=\lg X\rg.
$$
Since $F=\lg X(n,S)\rg$ is a free group on $X(n,S)$, there exists a uniquely
determined epimomorphism $\varphi:F\to H$ coinciding with $f^{-1}_\X$ on
$W_{X(n,S)}$ where $f_\X:W_\X\to W_{X(n.S)}$ is a bijection taking
$h_1\cdots h_k$ to $x_{h_1}\cdots x_{h_k}$.  After identifying $W_\R$ with
the subset of $W_\X$ we have $F=\varphi^{-1}(H)\supset
\lg f_\X(\X\cup R)\rg\supset\lg X\rg=G$. Thus
$G<F<\GLZ$ and the mapping
\qtnl{110303e}
f:G\to H,\quad g\mapsto\varphi(g)
\eqtn
is a homomorphism such that $f(x_hr_h)=\varphi(x_h)\varphi(r_h)=h\cdot 1=h$
for all $h\in\X$. Now we can define a homomorphic public-key cryptosystem
$\S(H,n,S)$ over the group~$H$ with respect to the homomorphism~(\ref{110303e})
as follows:

\vspace{4mm}
\noindent{\bf Public Key:} the subset $X=X(n,S,R)$ of $\GLZ$ where $R$ is a
random subset of $W_\R$, and a bijection $\X\to X,\ h\mapsto x_hr_h$.

\vspace{2mm}
\noindent{\bf Secret Key:} the pair $(n,S)$.

\vspace{2mm}
\noindent{\bf Encryption:} given a plaintext $h\in H$ encrypt as follows:

\begin{itemize}
\item[]{\bf Step 1.} If $h=h_1\cdots h_k$ with $h_i\in\X$ for all~$i$, set
$M_h=(x_{h_1}r_{h_1})\cdots(x_{h_k}r_{h_k})$.
\vspace{2mm}

\item[]{\bf Step 2.}
Find an $\X$-representation $w_r=h'_1\cdots h'_m$ of a random $r\in W_{\R}$.
Set $M_r=x_{h'_1}\cdots x_{h'_m}$.
\vspace{2mm}

\item[]{\bf Step 3.} Output the matrix $M_rM_h\in\GLZ$ as the ciphertext
of~$h$.
\end{itemize}
\vspace{2mm}

\noindent{\bf Decryption:} given a cyphertext $g\in G$ decrypt as follows.

\begin{itemize}
\item[]{\bf Step 1.} Find the $X(n,S)$-representation $w_g=g_1\cdots g_k$ of
the element~$g$ (Theorem~\ref{110303b}).
\vspace{2mm}

\item[]{\bf Step 2.} Output $f^{-1}_\X(g_1)\cdots f^{-1}_\X(g_k)$ as the
plaintext of~$g$.
\end{itemize}
\vspace{4mm}

The correctness of the encryption and decryption algorithms immediately
follows from the definitions. Moreover, by Theorem~\ref{110303b} the
decryption of the cryptosystem $\S(H,n,S)$ can be done within time
$(l(n)+l(S)+l(w_g)))^{O(1)}$.

\subsn{\bf Remarks on security of the cryptosystem $\S(H,n,S)$.}
First, we observe that the decryption problem, i.e. the problem of computing
$f(g)$ for an element $g\in G$, is polynomial-time reducible to the
representation problem $\P(\GLZ,X)$. Thus the difficulty of the direct way
to break $\S(H,n,S)$ is based on that of the special case of this
representation problem with the promise $X\subset G(n,S)$:

\prblm
Given a matrix belonging to a group $G\le G(n,S)$ find a short
$X$-representation of it under the assumption that such a representation
does exist.
\eprblm

One can make this problem even harder using for instance the Nielsen
transformations \cite{LS} to replace $X(n,S)$ by other set of generators not
necessarily being a basis of the group $G(n,S)$ (these transformations
consist in succesive replacing elements of generating set for their inverses
or products). A less direct way to break the cryptosystem $\S(H,n,S)$ could
consist in finding the number $n$ and the set $X$, in other
words, the secret key. This seems to be difficult.

Finally, it should be remarked that the cryptosystem $\S(H,n,S)$ can be
transformed to the homomorphic public-key cryptosystem in the sense
of~\cite{GP}. To do this it suffices to find a set $A$ and a trapdoor function
$P:A\to G$ such that $\im(P)=\ker(f)$, i.e. to get the exact sequence
$$
A\,\mapright{P}\,G\,\mapright{f}\,H\,\mapright{}\,\{1\}.
$$
However, this can be done by choosing $A$ to be the set $W_K$ where
$K=\{hh'(hh')^{-1}:\ h,h'\in H\}$, and $P=f_\X$
(we make use the fact that in this setting the group $H$ has to be small).
We do not dwell on details since we do not stick here with the definition
of~\cite{GP}.

\subsn\label{110303a}{\bf Proof of Theorem~\ref{110303b}.}
The proof of the theorem is based on lemmas~\ref{110303h} and~\ref{110303i}.
In the first
of them the free group $\FF$ on~$\X$ is considered as the subset of the set
$W_\X$: any element of~$\FF$ is an irreducible word of $W_\X$ and the
identity of~$\FF$ is the empty word $1_\X\in W_\X$. The length of
the $\X$-representation of an element $g\in\FF$ is denoted by $|g|$. For an
arbitrary word $w\in W_\X$ we denote by $\ov w$ the element of~$\FF$
corresponding to~$w$. Below we will use an observation
from the proof of~\cite[Proposition~3.1]{LS} that if $\X=\{A,B\}$ and
$S\subset\ZZ$ is a nonempty finite set, then the elements $A^{-s}BA^s$,
$s\in S$, form a basis of a free subgroup of the group~$\FF$.

\lmml{110303h}
Let $\FF$ be a free group of rank $2$ on $\X=\{A,B\}$ and $G$ be a
subgroup of $\FF$ generated by the set $X=\{A^{-s}BA^s:\ s\in S\}$ where
$S\subset\ZZ$ is a nonempty finite set. Then given an element $g\in\FF$ one
can test whether $g\in G$ or not in time $(l(g)+l(S))^{O(1)}$ where $l(g)$
is the bit size of the $\X$-representation of~$g$; moreover, if
$g\in G$, then the $X$-representation $w_g$ can be found within
the same time and $l(g)\leq 3l(w_g)l(S)$.
\elmm
\proof To prove the lemma let us consider the following algorithm which for a
given element $g\in\FF$ by recursion on the length $|g|$ of its
$\X$-representation produces a certain pair $(i_g,w_g)\in\{0,1\}\times W_X$
such that $g\in G$ if and only if $i_g=1$ and $w_g$ is the
$X$-representation of~$g$.

\begin{itemize}
\item[]{\bf Step 1.} If $g=1_\X$, then output $(1,1_X)$. Otherwise, let
$u=A^aB^bA^c\cdots$ for suitable $a,b,c,\ldots\in\ZZ$.
\vspace{2mm}

\item[]{\bf Step 2.} If either $-a\not\in S$ or $(-a,b)\in S\times\{0\}$,
then output $(0,1_X)$. Otherwise set $u=A^{a+c}\ldots$.
\vspace{2mm}

\item[]{\bf Step 3.} Recursively find $(i_h,w_h)$ where $h=\ov{u}$.
If $i_h=0$, then output $(i_h,w_h)$.
\vspace{2mm}

\item[]{\bf Step 4.} Output $(1,w_g)$ where $w_g=vw_h$ with $v=A^aB^bA^{-a}$.
\bull
\end{itemize}
\vspace{2mm}

We observe that each recursive call at Step~3 is applied to the element
$h\in\FF$ with  $|h|<|g|$, so the number of recursive calls is at most~$|g|$
and each step can be implemented in time $O(l(g)+l(S))$. Thus the running
time of the algorithm is $(l(g)+l(S))^{O(1)}$. Next, due to the obvious
inequality  $l(c)\leq l(a+c)+l(a)$ we have
\qtnl{l290703}
l(g)=
l(A^aB^bA^c\cdots)\leq 2l(a)+l(b)+l(A^{a+c}\ldots)= 2l(a)+l(b)+l(h).
\eqtn
Since $w_g=vw_h$ and $v=(A^aBA^{-a})^b$ we get that $l(w_g)=l(b)+l(w_h)$.
On the other hand, $l(h)\leq 3l(w_h)l(S)$ by the recursive hypothesis. Thus
from (\ref{l290703}) it follows that
$$
l(g)\leq 2l(a)+l(b)+3l(w_h)l(S)=2l(a)+l(b)+3(l(w_g)-l(b))l(S)
\leq 3l(w_g)l(S)
$$
(we use that $l(b)\ne 0$ and $\max\{l(a),l(b)\}\leq l(S)$). This proves
the required inequality $l(g)\leq 3l(w_g)l(S)$.

To verify the correctness of the algorithm we need to show first that
$g\in G$ if and only if $i_g=1$, and second that if $i_g=1$, then $w_g$ is
the $X$-representation of~$g$. Using induction on $|g|$ suppose
that $g\in G\setminus\{1_\X\}$. We observe that the first term of an
arbitrary irreducible word $w\in W_\X$ such that $w=\ov{w'}$ for some
$w'\in W_X$, is of the form $A^a$ where $-a\in S$. So the output of Step~2
is correct. Moreover, from the definition of~$v$ at Step~4 it follows that
$v\in X$ and so $g\in G$ iff $h\in G$. Besides, if the algorithm terminates
at Step~3 or~4, then $i_g=i_h$ and by the induction hypothesis $w_h$ is
the $X$-representation of~$h$ iff $i_h=1$. Thus the output at
Step~3 is correct and $w_g\in W_X$. Since obviously
$$
g=\ov{vu}=\ov{v\ov{u}}=\ov{v\ov{w_h}}=\ov{vw_h}=\ov{w_g},
$$
we conclude that $w_g$ at Step~4 is the $X$-representation of~$g$
and the output of this step is correct.\bull

In the next lemma we deal with the subgroup of $\GLZ$ generated by the
set $X_n=\{A_n,B_n\}$ (see~(\ref{120403b})). Since this group is a free
group on~$X_n$, any element $M$ of it has the uniquely determined
$X_n$-representation coinciding with the irreducible word belonging to
$W_{X_n}$.

\lmml{110303i}
Let $G=\lg X_n\rg$ for some $n\ge 2$. Then given matrix $M\in\GLZ$ belonging
to~$G$, the $X_n$-representation of~$M$ can be found in time
$(l(n)+l)^{O(1)}$ where $l$ is the bit size this representation.
\elmm
\proof The algorithm below is similar to the one in \cite{S73} which yields
a representation of a matrix with respect to a different (more standard in
the theory of modular groups) family of generator, also in~\cite{S73} one
can find the basic facts on the group $\SLZ$ used in the proof below.
We will employ the classical action of the group $\GLZ$ on the
projective line (the Riemannian sphere) $\CC^*=\CC\cup\{\infty\}$ by means of
linear fractional transformations
$$
z\mapsto Mz=(M_{11}z+M_{12})/(M_{21}z+M_{22})
$$
where $M=(M_{ij})$ is a matrix of $\GLZ$ (the kernel of this action is of
order~2 and equal the subgroup of all diagonal matrices of $\GLZ$; the
quotient group with respect to this subgroup is the projective group
$\PGL_2(\ZZ)$). We make use of the following key observation: if $n\ge 2$,
then any power $A^k$ of the matrix $A=A_n$ with nonzero $k\in\ZZ$ maps the
unit open disk $D\subset\CC$ centered at~0 {\it strictly} inside
$D^c=\CC^*-\overline D$, and reciprocately any power $B^k$ of the matrix
$B=B_n$ maps $D^c$ {\it strictly} inside~$D$. \footnote{This observation
entails that $G$ is the free group on $\{A,B\}$
(see~\cite[Proposition~12.2]{LS}).}
A straightforward computation shows that given $z\in D\cup D^c$
there could exist at most one integer $k=k(z)$ such that
$$
(z\in D^c\ \land\ A^kz\in D)\quad\lor\quad(z\in D\ \land\ B^kz\in D^c).
$$
Below we set $C(z)=A^k$ if $z\in D^c$, and $C(z)=B^k$ if $z\in D$,
provided that $k$ does exist. In the following algorithm we suppose that
$I$ is the identity matrix, and $z\in D$ and $z'\in D^c$ are arbitrary
fixed complex numbers of small sizes, say $z=1/2$ and $z'=2$.

\begin{itemize}
\item[]{\bf Step 1.} Set $(L,L'):=(M,M)$ and $(u,u'):=(1_{X_n},1_{X_n})$.
\vspace{2mm}

\item[]{\bf Step 2.} If $L=I$, then output $u$; if $L'=I$, then output $u'$.
\vspace{2mm}

\item[]{\bf Step 3.} Set $(u,u'):=(C^{-1}u,(C')^{-1}u')$ (in
$W_{X_n}\times W_{X_n}$),
and $(L,L'):=(CL,C'L')$ (in $\GLZ\times\GLZ$), where $C=C(Lz)$, $C'=C(L'z')$.
Go to Step~2.\bull
\end{itemize}
\vspace{2mm}

Let us prove that the above algorithm finds the $X_n$-representation
\qtnl{0202a}
M=A^{a_1}B^{b_1}\cdots A^{a_m}B^{b_m}
\eqtn
of a matrix $M\in G$ where $m$ is a nonnegative integer and
$a_i,b_i\in\ZZ$, $i\in\ov m$, such that $a_i\ne 0$ for $i\ne 1$, $b_i\ne 0$
for $i\ne m$. If $M=I$ ($m=0$), then the statement is obvious (see Step~1).
Let us show that if $b_m=0$ (resp. $b_m\ne 0$), then after $m$ iterations of
the loop at Steps~2 and~3 the matrix $L$ (resp. $L'$) becomes the identity
matrix and the word $u$ (resp. $u'$) is the $X_n$-representation of~$M$.
Indeed, let $b_m= 0$ (the case $b_m\ne 0$ is considered similarly). Then it
is easy to see that $Mz\in D$ iff $a_1=0$. So after the first iteration
according to Step~3 we have
$$
k(Mz)=\cases{-a_1, &if $Mz\in D^c$,\cr -b_1, &if $Mz\in D$,\cr}
$$
whence $u=A^{a_1}$ if $Mz\in D^c$ and $u=B^{b_1}$ if $Mz\in D$. Since
the number of factors in the $X_n$-representation of the matrix
$L$ after Step~3 equals $m-1$, the required statement follows by induction
on this number.

Let us estimate the running time of the algorithm. We observe that
from the previous paragraph it follows that the algorithm terminates after
$m$ iterations. So to complete the proof it suffices to note that the sizes
of all the intermediate matrices $L$ and $L'$ do not exceed
$O(ml(n)+l)$.\bull

Let us complete the proof of Theorem~\ref{110303b}. For an element
$g\in G(n,S)$ by means of Lemma~\ref{110303i} one can find first its
$X_n$-representation within time $(l(n)+l)^{O(1)}$ where $l=l(g)$ is
the bit-size of this representation. Subsequently applying
Lemma~\ref{110303h} one can find an $X(n,S)$-representation $w_g$ of $g$
within time $(l+l(S))^{O(1)}\leq(l(w_g)+l(S))^{O(1)}$.

\section{Homomorphic cryptosystems over finite rings}
\label{sec3}

Let $R$ be a finite commutative ring with identity and $G$ be a group. Then
it is easy to see that any homomorphism $\varphi:G\to R^\times$ where
$R^\times$ is the multiplicative group of~$R$, can be extended to the
homomorphism $\varphi':R[G]\to R[R^\times]$ of the group
rings taking $\sum_gr_gg$ to $\sum_gr_g\varphi(g)$. On the other hand, the
natural injection $R^\times\to R$ can be extended to the ring homomorphism
$\varphi'':R[R^\times]\to R$. We will say that the homomorphism
$f=\varphi'\circ\varphi''$,
\qtnl{100303c}
f:R[G]\to R,\quad \sum_gr_gg\mapsto\sum_gr_g\varphi(g)
\eqtn
is induced by the homomorphism~$\varphi$. From the computational point of
view the homomorphisms $\varphi$ and $f$ are closely related; more exactly
the problem of finding $\varphi(g)$ for $g\in G$ is polynomial time equivalent
to the problem of finding $f(g)$ for $g\in G$ (here we suppose the elements
of the group ring $R[G]$ are given by $R$-linear combinations of elements
of~$G$). This immediately implies the following statement.

\lmml{100303b}
Let $R$ be a finite commutative ring with identity such that there exists a
homomorphic public-key cryptosystem over the group $R^\times$ with respect to
an epimorphism $\varphi:G\to R^\times$ for some group~$G$. Then
one can design a homomorphic public-key cryptosystem over the ring $R$.
Moreover, the problems of breaking these two systems are polynomial-time
equivalent.\bull
\elmm

{\bf Proof of Theorem~\ref{100303a}.}
We recall that the ring $R$ being a commutative one is isomorphic to a direct
sum of local rings (see \cite{MD74}). If among these local rings there is at
least one not isomorphic to $\ZZ_2$ then the multiplicative group of this
ring is nontrivial and hence $|R^\times|\ne 1$. Thus by Lemma~\ref{100303b} it
suffices to find a homomorphic public-key cryptosystem over the group
$R^\times$. To do
this we observe that due to the commutativity of the ring $R$, we have
$R^\times=H_1\times\cdots\times H_k$ where $H_i$ is a cyclic group, $i\in[k]$.
So from \cite[Section~2]{GP} it follows that for each~$i$ there exists a
homomorphic public-key cryptosystem $\S_i$ over the group $H_i$ with respect
to an appropriate epimorphism $\varphi_i:G_i\to H_i$ with $G_i$ being a
finite Abelian group. Set $G=G_1\times\cdots\times G_k$ and $\varphi$ to be
the epimorphism $G\to H$ induced by the epimorphisms
$\varphi_1,\ldots,\varphi_k$.
Now, using cryptosystems $\S_i$, $i\in[k]$, one can form a homomorphic
public-key cryptosystem over the group~$R^\times$ with respect to the
epimorphism $\varphi:G\to R^\times$. Theorem is proved.\bull

Let $R$ and $\A$ are finite rings as in Theorem~\ref{100303a}. Then from
the proof of this theorem it follows that the size of $\A$  is
double exponential in the size of the ring~$R$. Indeed, $\A$ is the group
ring of the group~$G$ over~$R$, whence $|\A|=|G|^{|R|}$,
$|G|=|G_1|\cdots |G_k|$ and $|G_i|$ is exponential in $|H_i|$ (see
construction in~\cite[Section~2]{GP}). We will see below that under the
natural assumption on the presentation of~$\A$ it is difficult to reduce
the size of~$\A$ preserving the secrecy of the homomorphism $f:\A\to R$
(this extends the observation from~\cite{BY88}).

Let $\A$ be a finite ring of characteristic $m$ (i.e. the minimal integer which vanishes in $\A$)
and $\P(m)$ be the set of the highest prime powers dividing~$m$. Then it is easy to see that
\qtnl{100303d}
\A=\bigoplus_{q\in\P(m)}\A_q
\eqtn
where $\A_q=q'\A$ with $q'=m/q$, is an ideal of~$\A$ considered as a finite
ring of characteristic~$q$ with the identity $q'1$. For each $q$ the ring $\A_q$
is a linear space of the dimension $n_q=\log_p|\A_q|$ over the finite field
$\F_p$ of the prime order~$p$ dividing~$q$.
This implies that $\A$ can be considered as a subring of the matrix ring
$\Mat_n(\ZZ_m)$ where $n=\sum_qn_q$. To find a basis of a linear space could
be not easy a priori if a procedure of testing linear dependency is not known,
that is why the efficiency of embedding of $\A$ into a matrix ring depends on
the way how $\A$ is given. Now suppose that the size of $\A$ is at
most exponential in $|R|$. Then the dimension $n_q$  is
polynomial in $|R|$ and hence $n$, $\log m$ are less than $|R|^{O(1)}$.
In the following theorem we use a presentation of a ring homomorphism
which is analogous to the presentation of a group homomorphism from~\cite{GP}.

\thrml{100303e}
Let $R$ be a finite ring presented by the list of elements together with
the Cayley tables of its additive and multiplicative groups and $\A$ be a
subring of the ring $\Mat_n(\ZZ_m)$ where
$\max\{n,\log m\}\le|R|^{O(1)}$. Suppose that $f:\A\to R$ is a homomorphism
given by generators of the ideal $\ker(f)$, a transversal $X$ of $\ker(f)$
in $\A$ and the restriction of $f$ to $X$. Then given $a\in\A$ the element
$f(a)$ can be found in polynomial time in $|R|$.
\ethrm
\proof Using the decomposition (\ref{100303d}) one can reduce the problem of
computing $f(a)$, $a\in\A$, in polynomial time to $|\P(m)|$ problems of
computing $f_q(a_q)$, $q\in\P(m)$, where $a_q=aq'\in\A_q$ and $f_q:\A_q\to R_q$ is the
homomorphism induced by $f$. Thus without loss of generality we assume that
the characteristic of $\A$ equals $p^d$ for a prime~$p$ and $d\ge 1$. Since
$d\le\log m\le|R|^{O(1)}$ one can find an embedding $\A\to\Mat_{nd}(\ZZ_p)$
in time $|R|^{O(1)}$. Then the ideal $\ker(f)$ becomes a linear space over
a finite field $\F_p$ of dimension at most $(nd)^2$. Using  linear algebra
 over $\F_p$ a linear basis of this space can be found within
the same time. This enables us to solve efficiently whether or not an
arbitrary element $a\in\A$ belongs to $\ker(f)$.

Let now $a\in\A$. Then there exists the uniquely determined element
$x_a\in X$ such that $x_a-a\in\ker(f)$. Moreover, from the previous paragraph it
follows that this element can be found in time $|R|^{O(1)}$ (it suffices to
test for each $x\in X$ whether or not $x-a\in\ker(f)$). Since
$f(a)=f(a+x_a-a)=f(x_a)$ and the element $f(x_a)$ is known as the part of
presentation of~$f$, the element $f(a)$ can be found within the same
time.\bull

\end{document}